

\hfuzz=20pt
\magnification=1200
\hoffset=-.1in
\voffset=-.2in

\vsize=7.5in
\hsize=5.6in
\tolerance 10000

\baselineskip 12pt plus 1pt minus 1pt
\pageno=0
\centerline{\bf SELF-DUAL CHERN--SIMONS SOLITONS}
\smallskip
\centerline{{\bf IN (2~+~1)-DIMENSIONAL EINSTEIN GRAVITY}\footnote{*}{This
work is supported in part by funds
provided by the U. S. Department of Energy (D.O.E.) under contract
\#DE-AC02-76ER03069, the Ministry of Education and by the Science and
Engineering Foundation of Korea (CL),
and by the Swiss National Science Foundation (DC).}}
\vskip 24pt
\centerline{D.~Cangemi}
\vskip 12pt
\centerline{\it Center for Theoretical Physics}
\centerline{\it Laboratory for Nuclear Science}
\centerline{\it and Department of Physics}
\centerline{\it Massachusetts Institute of Technology}
\centerline{\it Cambridge, Massachusetts\ \ 02139\ \ \ U.S.A.}
\vskip 12pt
\centerline{and}
\vskip 12pt
\centerline{Choonkyu Lee}
\vskip 12pt
\centerline{\it Department of Physics and}
\centerline{\it Center for Theoretical Physics}
\centerline{\it Seoul National University}
\centerline{\it Seoul, 151--742, KOREA}
\vskip 1.5in
\centerline{Submitted to: {\it Physical Review D} (Brief Reports)}
\vfill
\noindent hepth@xxx/yymmnn

\noindent SNUTP\#92-25

\centerline{ Typeset in $\TeX$ by Roger L. Gilson}
\vskip -12pt
\noindent CTP\#2090\hfill April 1992
\eject
\baselineskip 24pt plus 2pt minus 2pt
\centerline{\bf ABSTRACT}
\medskip
We consider here a generalization of the Abelian Higgs model in curved space,
by adding a Chern--Simons term.  The static equations are self-dual provided
we choose a suitable potential.  The solutions give a self-dual
Maxwell--Chern--Simons soliton that possesses a mass and a spin.
\vfill
\eject
Einstein gravity in $2+1$ dimensions has attracted much attention recently,
not only as a theoretical laboratory for studying effects of quantized gravity
but also due to its direct physical relevance in cosmic string dynamics.$^1$
The $(2+1)$-dimensional Einstein gravity
is trivial in the absence of matter, while
introducing point$^{2,\,3}$ or line$^4$ sources alters the global geometry of
space-time in the following way.  The metric describing $N$ point particles
located at ${\bf r}_i$ ($i=1,\ldots,N$), with mass $M_i$ and spins $J_i$, is
known to have the form:$^5$
$$ds^2 = - \left( dt + G \sum^N_{i=1} J_i {\left( {\bf r} - {\bf
r}_i\right)\over \left| {\bf r} - {\bf r}_i\right|^2} \times d{\bf
r}\right)^2 + \left( \prod^N_{i=1} \left|{\bf r} - {\bf
r}_i\right|^{-2GM_i} \right) \left( d{\bf r}\right)^2\ \ .\eqno(1)$$
The spatial geometry is thus multi-conical, and with non-zero $J_i$ one finds
a time helical structure.$^2$

As has been known for some time, the non-singular counterparts for spinless
multi-particle systems are provided by static multi-vortex solutions of the
curved-space Abelian Higgs model in the self-dual limit.$^6$  One expects a
regular configuration of non-zero spin by introducing a Chern--Simons term.
Linet$^7$ introduced a non-self-dual model including this term.  With some
assumptions he shows that the asymptotic geometry is analogous to that of a
spinning particle. Recently, Valtancoli$^{8}$ considered a curved-space
self-dual model taking the Chern--Simons term as the entire gauge field
action.
We shall present here a model including both the Maxwell and the Chern--Simons
terms and show how it leads to self-dual equations.  The Higgs model and the
pure Chern--Simons model correspond to special limiting cases of our
treatment.

In flat-space it is known$^{9}$ that, with both Maxwell and Chern--Simons
terms, the simplest self-dual system is described by the action
$$\eqalign{I_{\rm flat} &= \int d^3x \Biggl\{
- {1\over4} F_{\mu\nu} F^{\mu\nu} +
{\kappa\over 4}\epsilon^{\mu\nu\lambda} F_{\mu\nu} A_\lambda - \left|
D_\mu\phi\right|^2 \cr
&\quad\qquad - {1\over 2} \left( \partial_\mu S\right)^2 - {1\over 2}
\left( e\left|\phi\right|^2 + \kappa S - ev^2\right)^2 - e^2 S^2 \left|
\phi\right|^2 \Biggr\} \ \ ,\cr}\eqno(2)$$
where $D_\mu = \partial_\mu -{\rm i}e
A_\mu$, $\phi$ is a complex scalar field and
$S$ a real one.  For a related curved-space self-dual system, it is then
natural to consider the following action
$$\eqalignno{I &= {1\over 4\pi G} \int d^3x \,\sqrt{-g}\, R + I_M\ \
,&(3\hbox{a}) \cr
I_M &= \int d^3x \,\sqrt{-g}\biggl\{ - {1\over 4} g^{\mu\rho} g^{\nu\sigma}
F_{\mu\nu} F_{\rho\sigma} + {\kappa\over 4} {1\over \sqrt{-g}}
\epsilon^{\mu\nu\rho} F_{\mu\nu} A_\rho - g^{\mu\nu} \left( D_\mu
\phi\right)^* D_\nu \phi\cr
&\qquad\quad - {1\over 2} g^{\mu\nu} \partial_\mu S \partial_\nu S - V (\phi,
S)\biggr\}\ \ ,&(3\hbox{b}) \cr}$$
where the precise form of the scalar potential $V(\phi,S)$ remains to be
fixed.  Our interest is in time-independent soliton-like configurations
satisfying corresponding matter field equations and Einstein's equations
$\left( R^{\mu\nu} - {1\over 2} g^{\mu\nu} R = 2 \pi G T^{\mu\nu}\right)$, and
we here want specifically a model for which the governing equations for the
solitons can be reduced to first-order self-dual (or Bogonol'nyi-type)
equations.

We may assume the general stationary metric
$$ds^2 = - N^2 \left( dt + K_i dx^i\right)^2 + \gamma_{ij} \, dx^i\, dx^j\ \
,\qquad (i,j=1,2) \eqno(4)$$
{\it viz.\/}
$$\eqalign{g_{00} &= - N^2\ \ ,\cr
g^{00} &= - {1\over N^2} + \gamma^{ij} K_i K_j\ \ ,\cr}\hskip .4in
\eqalign{g_{0i} &= - N^2 K_i\ \ ,\cr
g^{0i} &= - \gamma^{ij} K_j\phantom{\bigg|}\ \ ,\cr} \hskip .4in
\eqalign{g_{ij} &= \gamma_{ij} - N^2 K_i K_j \ \ ,\cr
g^{ij} &= \gamma^{ij}\phantom{\bigg|}\ \ .\cr}\eqno(5)$$
Here, $N\ge 0$, $K_i$ and $\gamma_{ij}$ are functions of ${\bf r} = (x^1, x^2)$
only.  The spatial metric $\gamma_{ij}$ will be used to move indices.

Instead of $A_i$ we find it convenient to use the fields
$$\overline{A}_i = A_i - K_i A_0\ \ ,\eqno(6)$$
so that $\overline{A}^i\equiv \gamma^{ij} \overline{A}_j = g^{i\mu} A_\mu$.
Then, denoting ${1\over
\sqrt{\gamma}} \epsilon^{ij} \partial_i \overline{A}_j = \overline{B}$,
${1\over \sqrt{\gamma}} \epsilon^{ij} \partial_i K_j=H$,
and
$\overline{D}_i=\partial_i-{\rm i}e\overline{A}_i$,
we obtain the following {\it
static\/} action from the action (3) with time-independent fields:
$$\eqalign{
I &= \int d^3x\,\sqrt{\gamma}\, N \Biggl\{ {1\over 4\pi G} \overline{R} +
{1\over 2} \left( - A^2_0 + {N^2\over 4\pi G}\right) H^2 - A_0 \overline{B} H
+ {1\over 2}\ {1\over N^2} \gamma^{ij} \partial_i A_0 \partial_j A_0 - {1\over
2} \overline{B}^2 \cr
&+\kappa{1\over N} A_0\overline{B} + {\kappa\over 2} {1\over N} A^2_0 H + e^2
{1\over N^2} A^2_0 |\phi|^2 - \gamma^{ij} \left( \overline{D}_i\phi\right)^*
\overline{D}_j \phi - {1\over 2} \gamma^{ij} \partial_i S \partial_j S -
V(\phi,S)\Biggr\}\ \ .\cr}\eqno(7)$$
Here $\overline{R}$ is the Ricci scalar associated with the metric
$\gamma_{ij}$.
The equation of motion coming from an $N$-variation give:
$$\eqalign{{1\over 4\pi G} \sqrt{\gamma}\,\overline{R} &=
\sqrt{\gamma}\Biggl\{ \left( {1\over 2} A^2_0 - {3N^2\over 8\pi G} \right) H^2
+
A_0 \overline{B}H + {1\over 2} {1\over N^2}
\gamma^{ij} \partial_i A_0 \partial_j A_0 + {1\over 2}
\overline{B}^2 + {1\over N^2} e^2 A^2_0 |\phi|^2 \cr
&\qquad\quad +\gamma^{ij} \left( \overline{D}_i\phi\right)^*
\overline{D}_j\phi
+ {1\over 2}
\gamma^{ij} \partial_i S \partial_j S + V(\phi,S)\Biggr\}\ \ .\cr}\eqno(8)$$
This is one of the Einstein's equations.  In what follows we will take as an
{\it Ansatz\/} that
$$N({\bf r}) = 1\ \ .\eqno(9)$$
This means puting $N=1$ in the action (7) and keeping Eq.~(8) as an extra
constraint that our solutions must fulfill.

The static field equation related to the $K_i$-variation of the action (7)
can readily be integrated to yield
$$\left( - A^2_0 + {1\over 4\pi G} \right) H - A_0\overline{B} + {\kappa\over
2} A^2_0 = - {\kappa\over 2} C\ \ , \eqno(10)$$
$C$ being an integration constant.  Then, making use of the identity
$$\eqalign{ \sqrt{\gamma}\, \gamma^{ij} \left( \overline{D}_i\phi\right)^*
\overline{D}_j \phi &= {1\over 2} \sqrt{\gamma}\,\gamma^{ij} \left[
\left( \overline{D}_i
\phi\right)^* \mp {\rm i} \sqrt{\gamma}\, \epsilon_{i\ell} \gamma^{\ell m}
\left(\overline{D}_m\phi\right)^* \right]
\left[ \overline{D}_j \phi \pm {\rm i} \sqrt{\gamma}\, \epsilon_{jk}
\gamma^{kn}
\overline{D}_n \phi\right] \cr
&\pm e\overline{B} |\phi|^2
\mp {{\rm i}\over 2}\ {1\over\sqrt{\gamma}} \epsilon^{ij}\partial_i \left(
\phi^* \overline{D}_j\phi - \left( \overline{D}_j\phi\right)^* \phi\right)
\cr} \eqno(11)$$
(note that $\epsilon_{ij} = \epsilon^{ij}$ in our convention), we observe that
the action (7) can be rearranged into the following form:
$$\eqalign{
I &= \int d^3x \sqrt{\gamma}\,\Biggl\{ {1\over 2} \left( - A^2_0 + {1\over
4\pi G}\right) \left[ H + {-A_0\overline{B} + {\displaystyle{\kappa\over 2}}
\left( A^2_0 + C\right) \over - A^2_0 + {\displaystyle{1\over 4\pi G}}}
\right]^2 \cr
&- {1\over 2}\ {1\over \left( 1 - 4\pi G A^2_0\right)}
 \biggl( \overline{B} -
2\pi G\kappa A_0 \left( A^2_0 +C\right) - \left( 1 - 4\pi G A^2_0\right)
\left[ \kappa A_0 \mp e\left( |\phi|^2 - v^2 \right) \right] \biggr)^2 \cr
&- {\pi G\kappa^2\over 2\left( 1 - 4\pi GA^2_0\right)} \left( A^2_0 +
C\right)^2 + {1\over 2} \left( 1 - 4\pi G A^2_0\right) \left[ {2\pi G \kappa
A_0 \left( A^2_0 + C\right) \over 1 - 4\pi G A^2_0} + \kappa A_0 \mp e\left(
|\phi|^2 - v^2 \right) \right]^2 \cr
&+ e^2 A^2_0 |\phi|^2 - V(\phi,S) + {1\over 2} \gamma^{ij} \partial_i A_0
\partial_j A_0 - {1\over 2}\gamma^{ij} \partial_i S \partial_j S \cr
&- {1\over 2} \gamma^{ij} \left[ \left( \overline{D}_i \phi\right)^* \mp{\rm i}
 \sqrt{\gamma}\,\epsilon_{i\ell}\gamma^{\ell m} \left(
\overline{D}_m\phi\right)^*
\right] \left[ \overline{D}_j\phi \pm {\rm i}
\sqrt{\gamma}\, \epsilon_{jk} \gamma^{kn}
\overline{D}_n\phi\right] \Biggr\}\ \ .\cr}\eqno(12)$$
Equation~(10) is manifestly incorporated in this form, and see below for the
role for an arbitrary constant $v^2$.  Now, as a natural extension of the
flat-space self-duality equations,$^{9}$ let us suppose that the following
equations hold:
$$\eqalignno{\overline{D}_i \phi &\pm {\rm i} \sqrt{\gamma}\,\epsilon_{ij}
\gamma^{jk} \overline{D}_k \phi = 0\ \ ,&(13\hbox{a})\cr
A_0 &= \mp S\ \ .&(13\hbox{b})\cr}$$
Then the field equation related to the $A_i$-variation will effectively take
the form
$$\overline{B} = 2\pi G \kappa A_0\left( A^2_0 + C\right) + \left( 1 - 4\pi G
A^2_0 \right) \left[ \kappa A_0\mp e \left( |\phi|^2 - v^2\right)\right] \ \
,\eqno(14)$$
with $v^2$ interpreted as the associated integration constant.

We still have to consider the field equations related to the $\gamma_{ij}$-,
$\phi$-, $S$- and $A_0$-variations of the action (12), and for these the
first, second and last terms in the right-hand side of the action can be
ignored thanks to Eq.~(10), (13) and (14).  The equations from the
$\gamma_{ij}$-variation then forces the remaining terms in the action (12) to
vanish locally when $A_0 = \mp S$; this fixes the scalar potential for a
consistent self-dual system to take the following sixth-order form
$$V(\phi,S) = {1\over 2} \left( e |\phi|^2 + kS - ev^2\right)^2 + e^2 S^2
|\phi|^2 - 2\pi G \left( {1\over 2}\kappa (S^2 - C) + \left( e |\phi|^2 -
ev^2\right) S\right)^2 \ \ .\eqno(15)$$
On the other hand, it follows that the only non-trivial equation obtained from
the $\phi$-, $S$- and $A_0$-variations is (here, $\overline{\nabla}$ is the
two-dimensional covariant derivative)
$$\overline{\nabla}^i \overline{\nabla}_i S - {\partial V(\phi,S)\over
\partial S} = 0\ \ ,\eqno(16)$$
which, as an equation for $A_0$, describes Gauss' law.  We can now state that
if the scalar potential of a given system is the one given in Eq.~(15), any
static configuration satisfying the conditions (8), (9), (10), (13), (14) and
(16) --- the desired curved space generalization of the flat-space
self-duality equations --- provides a solution to the full coupled field
equations. Also notice that Eqs.~(10) and (14) are equivalent to the relations
$$\eqalign{ H &= 4\pi G \left[ - {\kappa\over 2} \left( S^2 + C\right) +
S\left( e|\phi|^2 + \kappa S - ev^2\right) \right]\ \ ,\cr
\overline{B} &= \mp \left[ \left( e |\phi|^2 + \kappa S - ev^2\right) -
SH\right]\ \ ,\cr}\eqno(17)$$
while Eq.~(8) can be simplified (thanks to other equations in the set) as
$${1\over 4\pi G} \overline{R} = \pm e v^2 \overline{B} + \kappa CH + {1\over
2} \sqrt{\gamma}\, \overline{\nabla}^i \overline{\nabla}_i |\phi|^2 +
{1\over 2} \sqrt{\gamma}\, \overline{\nabla}^i \overline{\nabla}_i S^2\ \
.\eqno(18)$$

Actually, the integration constant $C$ above is subject to a physical
constraint: at spatial infinity, the fields $(S,\phi)$ should approach some
constant values $(S_\infty,\phi_\infty$) and $H$ and $\overline{B}$ should
tend to zero (since their integrated values correspond to physical observables
to be discussed below).  Then we have from Eq.~(17)
$$e\left| \phi_\infty\right|^2 + \kappa S_\infty - ev^2 = 0\ \ ,\qquad
S^2_\infty + C = 0 \ \ ,\eqno(19)$$
while Eq.~(16) requires in addition
$$S_\infty \left| \phi_\infty\right|^2 = 0\ \ .\eqno(20)$$
Thus the allowed values for $C$ are
$$C = \cases{ 0 & (with $|\phi_\infty|=v$ and $S_\infty=0$) \qquad
\cr\noalign{\vskip 0.2cm}
-{\displaystyle{e^2v^4\over \kappa^2}}
 & (with $|\phi_\infty|=0$ and $S_\infty={\displaystyle{ev^2\over
\kappa}}$)\ \ .\cr}\eqno(21)$$
With $C=0$ we have the broken vacuum and topological soliton solutions, while
with $C = -e^2v^4/\kappa^2$ we have the unbroken vacuum and non-topological
solitons only.  (But, if we turned off gravity ({\it i.e.\/} set $G=0$), both
would of course lead to a theory in which broken and unbroken vacua are
degenerate.$^{9}$)  Regardless of $C=0$ or $C= -e^2v^4/\kappa^2$, the total
energy$^{2,\,10}$ of the given soliton is
$$\eqalign{E &= {1\over 4\pi G}\int d^2{\bf r}\,\sqrt{\gamma}\,
\overline{R} \cr
&= \pm ev^2\Phi\ \ ,\qquad \left( \Phi\equiv \int d^2{\bf r}\, \epsilon^{ij}
\partial_iA_j\right)\cr}\eqno(22)$$
where we have used Eq.~(18) and the relation $A_0 = \mp S$.  As in the
flat-space
case, the magnetic flux of a topological soliton must be quantized, {\it
i.e.\/} $\Phi = \pm {2\pi\over e} n$ ($n$: positive integer) while the
$\Phi$-value of a non-topological soliton is not.  These solitons have also
non-zero angular momentum, as determined by the formula$^{2,\,10}$
$$\eqalign{ J &= {1\over 2\pi G}\int d^2{\bf r}\, \sqrt{\gamma}\, H \cr
&= \int d^2{\bf r}\,
\sqrt{\gamma}\,\biggl\{ - \kappa \left( S^2+C\right) + 2S \left(
|\phi|^2 + \kappa S - ev^2\right)\biggr\}\ \ .\cr}\eqno(23)$$
Note that this definition is consistent with the asymptotic form $K_i({\bf
r})\sim - GJ\epsilon_{ij} {x^j/ |{\bf r}|^2}$ [{\it cf}.~Eq.~(1) and (4)], and
for $G=0$ it reduces to the usual flat-space angular momentum appropriate to
the topological ($C=0$) or non-topological ($C=-e^2v^4/\kappa^2$) soliton
case.

We here mention some limiting cases.  When the Chern--Simons coupling $\kappa$
is equal to zero, it is consistent to set $A_0 = S = K_i = 0$ and our system
trivially reduces to the model considered in Ref.~[6].  The scalar potential
becomes simply $V (\phi) = {1\over 2} e^2 \left( |\phi|^2 - v^2\right)^2$, and
we here have the self-duality equations:
$$D_i\phi \pm {\rm i} \sqrt{\gamma}\, \epsilon_{ij} \gamma^{jk} D_k\phi = 0 \ \
,\qquad {1\over \sqrt{\gamma}} \epsilon^{ij} \partial_i A_j = \mp e \left(
|\phi|^2 - v^2\right)\ \ .\eqno(24)$$
Only topological solitons are possible with $\kappa=0$.
More interesting will be
the limit $\kappa\to\infty$ for fixed $\kappa/e^2$.  In this limit, the
kinetic term for $S$ and the Maxwell term become negligible and up to
order-$(1/\kappa)$ corrections one can identify:
$$A_0  = \mp S = \pm {e\over\kappa}\left( |\phi|^2 - v^2\right)\ \ .\eqno(25)$$
Then one finds that the appropriate matter action and self-duality equations
read
$$\eqalignno{
I_M &= \int d^3x\Biggl\{ {\kappa\over 4} \epsilon^{\mu\nu\lambda}
F_{\mu\nu} A_\lambda - \sqrt{-g}\, g^{\mu\nu} \left( D_\mu \phi\right)^* D_\nu
\phi\cr
&\qquad\quad - \sqrt{-g}\left( {e^4\over \kappa^2} |\phi|^2 \left( |\phi|^2 -
v^2\right)^2 - {1\over 2} \pi G {e^4\over \kappa^2} \left[ \left( |\phi|^2 -
v^2\right)^2 + {\kappa^2\over e^2} C\right]^2\right)\Biggr\}\ \ ,
&(26)\cr\noalign{\vskip 0.2cm}
\overline{D}_i \phi&\pm {\rm i}\sqrt{\gamma}\,\epsilon_{ij} \gamma^{jk}
\overline{D}_k \phi = 0 \ \ ,&(27\hbox{a}) \cr\noalign{\vskip 0.2cm}
H &= - 2\pi G {e^2\over \kappa}\left[ \left( |\phi|^2 - v^2\right)^2 +
{\kappa^2\over e^2} C\right]\ \ ,&(27\hbox{b}) \cr\noalign{\vskip 0.2cm}
\overline{B} &= \mp {2e^3\over \kappa^2} |\phi|^2 \left( |\phi|^2 - v^2\right)
\pm 2\pi G {e^3\over \kappa^2} \left( |\phi|^2- v^2\right) \left[ \left(
|\phi|^2 - v^2\right)^2 + {\kappa^2\over e^2}C\right]\ \ , &(27\hbox{c})\cr}$$
where $C=0$ for the broken phase case (topological soliton solutions only) and
$C = - e^2v^4/\kappa^2$ for the unbroken phase case (non-topological soliton
solutions only).  Note that we have now a eight-order potential.
The total energy is still given by Eq.~(22) while the
angular momentum formula simplifies as
$$J = - {e^2\over \kappa} \int d^2{\bf r}\,
\sqrt{\gamma}\left[ \left( |\phi|^2 -
v^2\right)^2 + {\kappa^2\over e^2}C\right]\ \ .\eqno(28)$$
The action (26), with $C$ set to zero, was first obtained by
Valtancoli.$^{8}$  (But this paper contains a few sign mistakes.)

To analyze the curved-space self-duality equations, particularly convenient is
the conformal coordinate system in which $\gamma_{ij} = \rho ({\bf r})
\delta_{ij}$ and so
$$\sqrt{\gamma} \, \overline{R} = - \Delta\ln\rho\ \ ,\qquad \sqrt{\gamma}
\, \overline{\nabla}^i \overline{\nabla}_i |\phi|^2 = \Delta |\phi|^2\ \ ,
\eqno(29)$$
where $\Delta$ is the flat-space Laplacian.  Moreover, to remove the
arbitrariness in $K_i$ associated with the time reparametrization $t\to t' = t
+ \Lambda({\bf r})$, we adopt the gauge condition
$$\overline{\nabla}_i K^i = 0\ \ .\eqno(30)$$
In the conformal coordinates, this
condition is equivalent to $\partial_i K_i = 0$ and therefore we may write
$$K_i = \epsilon_{ij} \partial_j U({\bf r}) \ \ ,\qquad \sqrt{\gamma}\, H =
- \Delta U({\bf r}) \ \ .\eqno(31)$$
Similarly, we may express the vector potential $\overline{A}_i$ as
$$\overline{A}_i = - {1\over 2} \epsilon_{ij} \partial_j \overline{\cal A}
({\bf r})\ \ ,\qquad \sqrt{\gamma} \,\overline{B} = {1\over e} \Delta
\overline{\cal A}({\bf r}) \ \ .\eqno(32)$$
Using these in Eqs.~(13a) and (18) and choosing in particular the value $C=0$,
we can solve the equations for $\phi({\bf r})$ and $\rho({\bf r})$, to
obtain:
$$\eqalignno{\phi({\bf r}) &= e^{\mp\overline{\cal A}({\bf
r})}f(z)\ \ ,\qquad (z\equiv x\pm iy) &(33\hbox{a}) \cr
\rho({\bf r}) &= \left( {\left| \phi({\bf r})\right|^2 \over \left|
f(z)\right|^2}\right)^{2\pi G v^2} e^{ - 2\pi G \left[ \left|
\phi({\bf r})\right|^2 + S({\bf r})^2 \right]}\ \ ,&(33\hbox{b})
\cr}$$
where $f(z)$ can be any finite polynomial in $z$.  Here note that Eqs.~(32)
and (33a) allow us to write
$$\rho({\bf r})  \overline{B} ({\bf r}) = \mp {1\over 2e}
\Delta \left[ \ln
{\left|\phi({\bf r})\right|^2\over \left| f(z)\right|^2}\right]\ \
.\eqno(34)$$
Now, if we use Eqs.~(33b) and (34) in Eq.~(14), we are left with the equation
$$\Delta\ln |\phi|^2 =2e \left( {|\phi|^2\over |f(z)|^2} \right)^{2\pi Gv^2}
e^{-2\pi G\left[ |\phi|^2 + S^2\right]} \biggl[ \left( 1 - 4\pi
GS^2\right) \left( e|\phi|^2 + \kappa S - ev^2\right) + 2\pi G\kappa
S^3\biggr]\ \ ,\eqno(35)$$
which is valid away from the zeroes of $|\phi|^2$.  In this way we can reduce
the whole problem to the analysis of the two coupled equations involving
$|\phi|^2$ and $S$, {\it i.e.\/} Eqs.~(35) and (16).  In the $\kappa\to\infty$
limit mentioned earlier ({\it i.e.\/} for the system described by the action
(26)), they become just one non-trivial equation:
$$\Delta\ln |\phi|^2 = \left( {|\phi|^2\over |f(z)|^2}\right)^{2\pi G v^2}
e^{-2\pi G |\phi|^2} \left\{ {4e^4\over \kappa^2} |\phi|^2 \left( |\phi|^2-
v^2\right) - 4\pi G {e^4\over \kappa^2} \left( |\phi|^2 - v^2
\right)^3\right\}\ \ .\eqno(36)$$
But, even for the latter case, some numerical analysis appears to be necessary
for more detailed information.  For discussions on the asymptotic behaviors of
the solutions, see Ref.~[8].

Finally, we would like to add some comments on the stability of our system.  A
specific concern here is that, since our scalar potential (15) is unbounded
from below (for $G>0$), the vacua assumed in Eq.~(21) will be at
most local minima classically.  But it must be noted that, in the presence
of gravity, the definition of energy depends on the asymptotic behaviors of
the metric and so there is no simple way to compare the energies of two
different vacua.  Furthermore, self-dual systems are generally believed to be
the bosonic sector of some extended supersymmetric theories$^{11,\,12}$ and we
naturally expect our present system to be related to a certain extended
supergravity theory.  In the latter framework, the vacuum stability is likely
to follow automatically.$^{10}$
\goodbreak
\bigskip
\centerline{\bf ACKNOWLEDGMENTS}
\medskip
\nobreak
This work was undertaken when one of us (CL) was visiting the Center for
Theoretical Physics, MIT (as a part of the NSF--KOSEF Exchange Program), and
he wishes to thank members of the Center for hospitality.  We thank R.~Jackiw
for helpful comments.
\vfill
\eject
\centerline{\bf REFERENCES }
\medskip
\nobreak
\item{1.}A. Vilenkin, {\it Phys. Rev. D\/} {\bf 23}, 852 (1981); J. Gott, {\it
Ap. J.\/} {\bf 288}, 422 (1985).
\medskip
\item{2.}S. Deser, R. Jackiw and G. 't~Hooft, {\it Ann. Phys.\/} (NY) {\bf
152}, 220 (1984); A. Staruszkiewicz, {\it Acta. Phys. Polon.\/} {\bf 24}, 734
(1963); J. Gott and M. Alpert, {\it Gen. Rel. Grav.\/} {\bf 16}, 243 (1984);
S. Giddings, J. Abbot and K. Kuchar, {\it Gen. Rel. Grav.\/} {\bf 16}, 751
(1984).
\medskip
\item{3.}For a recent review, see R. Jackiw, ``Five Lectures on Planar
Gravity,'' SILARG VII, Cocoyoc, Mexico (December 1990), MIT preprint
CTP\#1936; and ``Update on Planar Gravity,'' Marcell Grossmann
Meeting, Kyoto, Japan (June 1991), MIT preprint CTP\#1986.
\medskip
\item{4.}S. Deser and R. Jackiw, {\it Ann. Phys.\/} (NY) {\bf 192}, 352
(1989); G. Grignani and C. Lee, {\it ibid.\/}, {\bf 196}, 386 (1989); G.
Cl\'ement, {\it ibid.\/} {\bf 201}, 241 (1990).
\medskip
\item{5.}G. Cl\'ement, {\it Int. J. Theor. Phys.\/} {\bf 24}, 267 (1985).
\medskip
\item{6.}B. Linet, {\it Gen. Rel. and Grav.\/} {\bf 20}, 451 (1988); A. Comtet
and G. W. Gibbons, {\it Nucl. Phys.\/} {\bf B299}, 719 (1988).
\medskip
\item{7.}B. Linet, {\it Gen. Rel. and Grav.\/} {\bf 22}, 469 (1990).
\medskip
\item{8.}P. Valtancoli, Universit\`a di Firenze preprint (July 1991).
\medskip
\item{9.}C. Lee, K. Lee and H. Min, {\it Phys. Lett.\/} {\bf 252B}, 79
(1990).
\medskip
\item{10.}M. Henneaux, {\it Phys. Rev. D\/} {\bf 29}, 2766 (1984).
\medskip
\item{11.}E. Witten and D. Olive, {\it Phys. Lett.\/} {\bf 78B}, 97 (1978); P.
DiVecchia and S. Ferrara, {\it Nucl. Phys.\/} {\bf B130}, 93 (1977).
\medskip
\item{12.}C. Lee, K. Lee and E. Weinberg, {\it Phys. Lett.\/} {\bf 243B}, 105
(1990); B.-H. Lee, C. Lee and H. Min, Seoul National University preprint
(1992).
\par
\vfill
\end